# Scalable, low-cost, and versatile system design for air pollution and traffic density monitoring and analysis

Thinh G. Tran[1], Dat T. Vo[2], Long C. Tran[3], Hoang V. Pham[1], Chuong D. Le[1], An D. Le[4], Duy A. Pham[5] and Hien B. Vo[1]

[1] Vietnamese-German University, Binh Duong, Vietnam
[2] University of Windsor, Windsor, Canada
[3] Aalto University, Helsinki, Finland
[4] University of California San Diego, California, USA
[5] Computer Science Department, Hochschule Bonn-Rhein-Sieg, Sankt Augustin, Germany
giathinhtran3@gmail.com

**Abstract.** Vietnam requires a sustainable urbanization, for which city sensing is used in planning and decision-making. Large cities need portable, scalable, and inexpensive digital technology for this purpose. End-to-end air quality monitoring companies such as AirVisual and Plume Air have shown their reliability with portable devices outfitted with superior air sensors. They are pricey, yet homeowners use them to get local air data without evaluating the causal effect. Our air quality inspection system is scalable, reasonably priced, and flexible. Minicomputer of the system remotely monitors PMS7003 and BME280 sensor data through a microcontroller processor. The 5-megapixel camera module enables researchers to infer the causal relationship between traffic intensity and dust concentration. The design enables inexpensive, commercial-grade hardware, with Azure Blob storing air pollution data and surrounding-area imagery and preventing the system from physically expanding. In addition, by including an air channel that replenishes and distributes temperature, the design improves ventilation and safeguards electrical components. The gadget allows for the analysis of the correlation between traffic and air quality data, which might aid in the establishment of sustainable urban development plans and policies.

## 1 Introduction

The rise of the population is a serious concern that people discuss worldwide. As a consequence, the air environment in urban areas is getting more polluted. The government and researchers need air data from the city to find the root of the pollution and come up with a solution. To capture this, we come up with our instrument. The target is to capture the particulate matter (PM 1, PM 2.5, and PM 10), temperature, humidity and atmospheric pressure. We also aim to record the video simultaneously to study and find the relation between the number of vehicles and air quality and use that for future prediction.

At the early stage of the project, a dust measuring device that contains a STM32 microcontroller with PMS7003 for dust measuring and BME280 for temperature, humidity and pressure capture is used with a surveillance camera for each node. This combination started to show the weakness of storing the data. There are two main problems towards it: (1) the data name format; the commercial camera fixes the name file format, and adding other information, such as location, is impossible; (2) the video coming from the surveillance camera is saved in the AVI format to save spacing storage, and to use in the visual processing model is time-consuming due to the conversion of the file format.

With the help of Raspberry Pi 3B + as a mini-computer, all of the data files can be set to the desired output. Then the STM32 controller board with sensors is redesigned to become a module using the serial connection. A 5 Mpx camera module is used to capture visual data. We can upgrade or swap the broken modules by designing all the components in a modular form. The Raspberry Pi 3B+ connects to Microsoft Azure Blob, which provides storage for all data, and this process is automated. Azure Blob is flexible in storing the data, switching between archive and cool storage based on regularity to save unnecessary cost.

## 2 Related work

Regarding the work, a research in Thailand also has a similar approach [1] in which their main goal was to record PM2.5, PM10, temperature, humidity along with vehicle flow, categorized into five groups: light duty gasoline vehicles, light duty liquid petroleum gas (LPG) vehicles, heavy-duty diesel vehicles, motorcycles, and light duty diesel vehicles. To capture particulate data, a PTFE filter was used with an air pump, then weight the collected matter on an ultra-microbalance with the lowest readability of 0.1 μg. Temperature and humidity were recorded on a separate instrument for the meteorological data. Finally, the traffic flow was recorded with a closed-circuit camera. They have recognized the correlation between the traffic flow and PM is a minor factor in indoor air quality, but temperature and relative humidity seem to have a strong influence. However, they wrote that the limited monitored area could lead to inaccurate results.

In addition, Zhang et al. in the US developed a system for indoor air quality monitors using Raspberry Pi as the central controller [2]. They designed a system that can log SO2, NO2, TVOC, O3, CO, CO2, PM2.5, PM10, temperature, and humidity. They also simplified the sensor and Raspberry Pi connection using a USB connector. However, their Low-Cost Air Quality Sensors system is designed for interior monitoring and cannot thus survive outdoor conditions.

In Europe, the We-count project aims to measure mobility and air quality data using Raspberry Pi[3]. They have launched the pilot tests in Madrid, Ljubljana, Dublin, Cardiff, and Leuven. From these nodes, the project can provide a data set of vehicle and air quality data for the scientists to research air pollution correlation with traffic. The We-count project includes two devices: a Raspberry Pi camera system and an IoT gadget for measuring PM. Volunteers install the visible data logging system on an interior glass window, while the dust data logging equipment is erected outdoors beneath the window roof. Two distinct gadgets might complicate the mapping network.

## 3 Methodology

Our system aims to gather video traffic data and environmental parameters such as dust concentration, temperature, humidity and atmospheric pressure. All data is stored on the cloud, through the Azure Blob Storage service. We aim to make it low-cost and scalable simultaneously.

Our system consists of 3 hardware components (Fig. 1): 1 Raspberry Pi mini-computer, 1 onboard camera, namely Pi Camera V2, and 1 Environmental Monitoring Module, which is custom built by us to collect environmental parameters through a series of sensors.

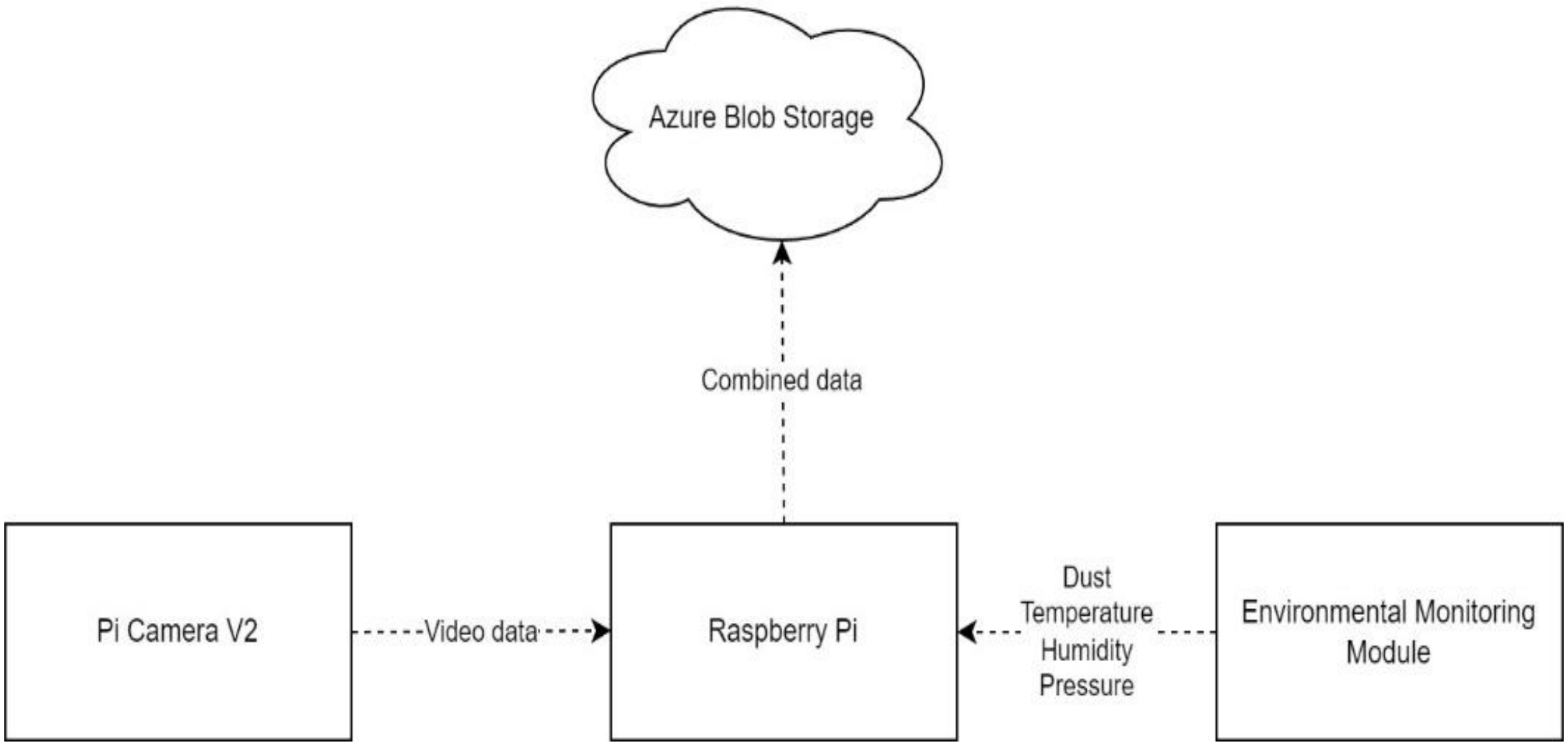

**Fig. 1.** Block diagram illustrating the components of the system

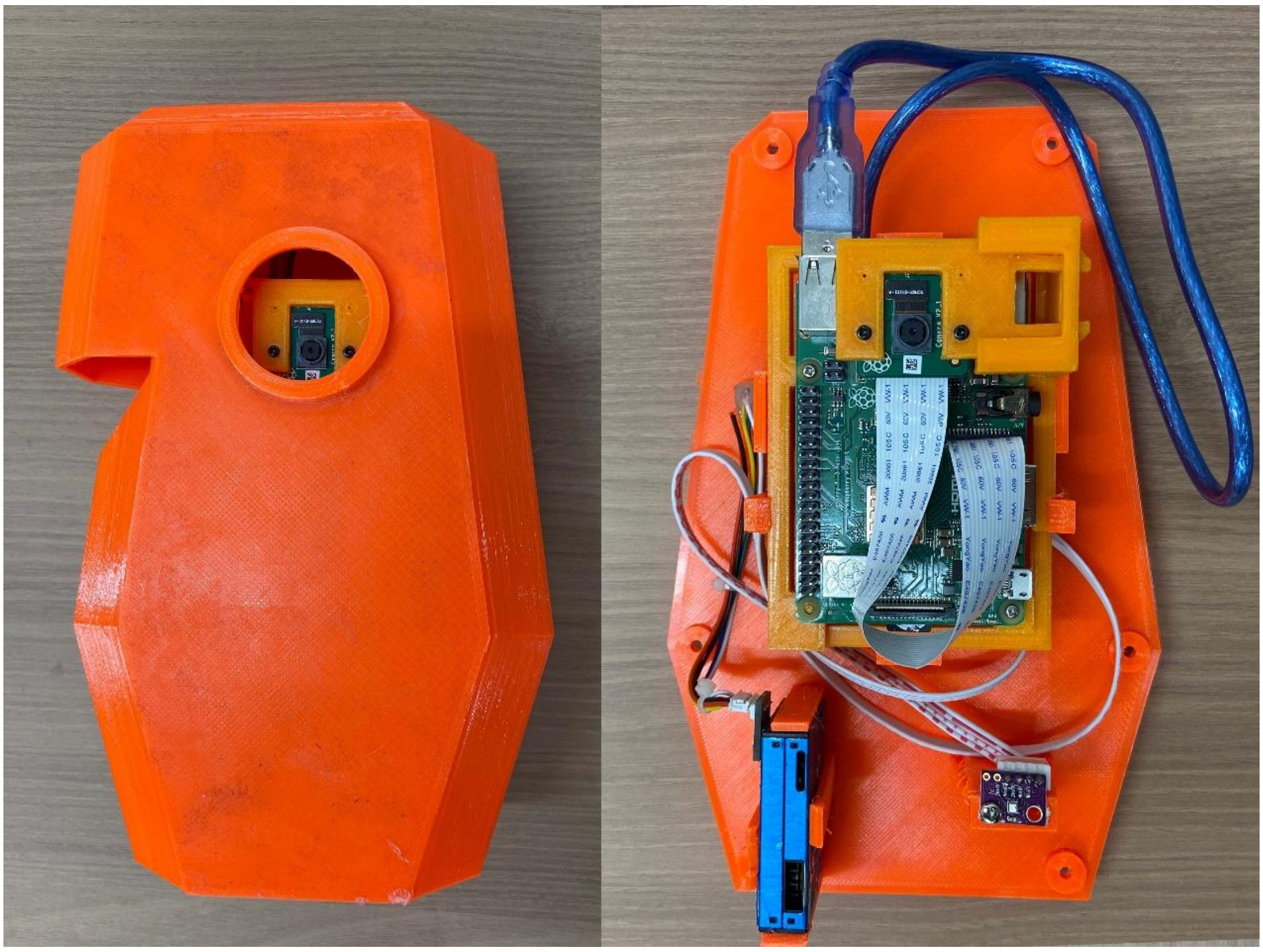

**Fig. 2.** The system's enclosure and internal circuitry

### 3.1 Environmental Monitoring Module

This module aims to gather data on environmental parameters through a series of sensors, then process them and export the complete data through its USB port. This module is then connected through a USB cable to the Raspberry Pi 3B+, where the data will be stored into a CSV file and uploaded daily to the cloud. Data is sampled and sent out at a 10-second interval. The whole module is mounted beneath the Raspberry Pi.

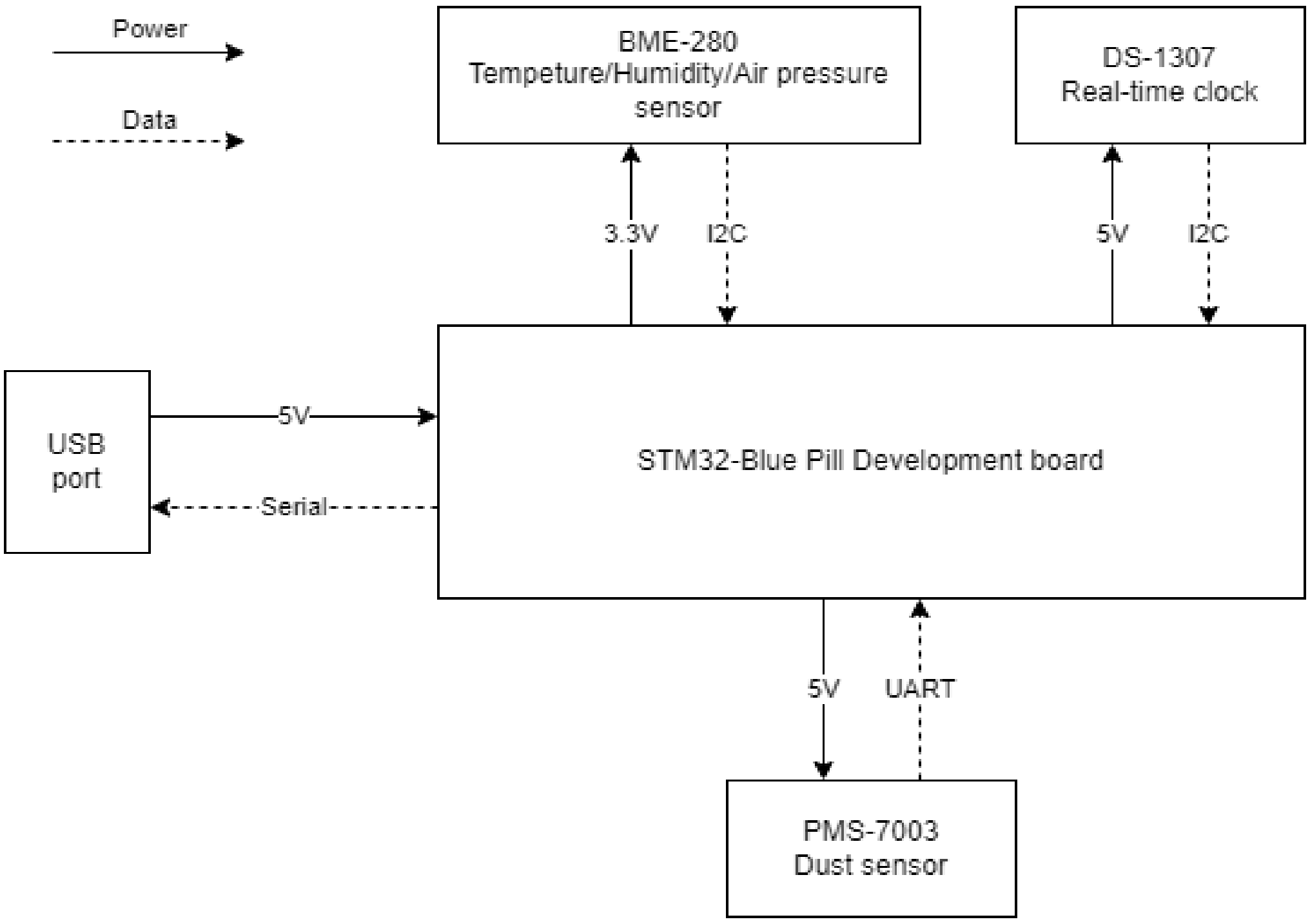

**Fig. 3.** The block diagram illustrating the components of the environmental monitoring module

**STM-32 Blue Pill development board.** The STM-32 Blue Pill was explicitly chosen for its abundance of connection ports. It supports 37 GPIO pins, 3 UART, 2 I2C and 2 SPI, 1 USB 2.0 as well as 1 CAN 2.0 peripherals [4]. In our particular application, we only utilize 2 UART and 1 I2C interface. But if we want to make additional changes to the module in the future, such as adding a CO2 or an insert gas sensor, we can very easily use the remaining ports. The high processing speed and large memory capacity also contribute to our choice of the particular microcontroller.

**BME-280 sensor.** The BME-280 sensor is a module used to record temperature, humidity and atmospheric pressure data. This sensor is chosen because of its compact build and high accuracy.

**DS-1307 Real-time clock.** The clock is used to reliably keep track of the time. Raspberry Pi doesn't have an onboard real-time clock, so we need an external clock to ensure correct timekeeping.

**PMS-7003 Dust sensor.** The sensor is used to monitor the concentration of particles of different sizes, namely PM1.0, PM2.5 and PM10. This sensor is chosen because of its low price and high accuracy. PMS-7003 when compared to its peers shows a better general performance, with low intra-model variability, high linear fitting coefficient with superior sensor and high durability in adverse conditions. [5]

### 3.2 Raspberry Pi and onboard camera

The goal of this project is to capture video data of traffic and send it to the cloud for further processing. To achieve that, we utilize a Raspberry Pi 3B+ and an onboard camera.

**The camera.** The camera chosen for this project is the Pi Camera V2. It has an 8 Megapixel sensor, which provides 1080p resolution at 30 frames per second. It has a field of view of 62 degrees [6].

To reduce file size and save storage space, we only set the camera to operate at 10 frames per second, 1296x730 resolution. We determined that 10 frames per second is sufficient for our application.

**The Raspberry pi.** The Raspberry Pi 3B+ was chosen because it can record videos locally and upload them to cloud storage through API. Another approach was to use Microcontroller Units with Wifi capabilities such as the ESP32. However, because we plan to implement this as an edge device in the future, we decided to use Raspberry Pi.

The Pi operates in headless mode and continuously runs a Python script. It will record videos and cut them into 5 minutes files. For air quality data, it will receive data from the Environmental Monitoring Module, parse it, and store it in a CSV file. These files will be uploaded to our Azure Blob Storage, the method of which will be discussed in later sections.

### 3.3 Storage solution

In order to have high scalability, we need to implement cloud data storage. The service we chose to use is Microsoft Azure Blob Storage which provides low-cost, high-volume data storage.

We assign each node in our system to 1 blob. Each time a new node is set up, it will be automatically assigned to a blob bearing its name. All video and text data of 1 node is uploaded into the same blob. We used Azure's recommended Python library to upload the data.

First, data is collected and stored locally in a file. For video, each file is 5 minutes long. For text data, each file contains the data for 24 hours. Second, the files are read as a byte stream, and then this byte stream is directed to the pre-specified blob. The upload process lasts around 5-7 minutes on a good Internet connection for a typical 5-minute-long video. Due to this reason, each upload operation has to be carried out asynchronously on a separate sub-process, to avoid blocking the main loop. Lastly, after completing the upload process, the files stored in the local drive will be deleted after a period of time specified by the setting. In this sense, the local storage acts as buffer storage.

It is estimated that one node will generate approximately 30 GB worth of video data daily. The cost for storing such amount on Azure Blob Storage will depend on the storage location, pricing plan and other settings [7]. However, because the daily fee will increase linearly over time when data accumulates, the total cost will increase exponentially over time. Therefore, additional operations to move unused data into archive storage are highly recommended for any attempt to replicate this method. However, for the scope of this project, we didn't implement such operations.

### 3.4 Enclosure

The current enclosure design is for quick installation while minimizing the needed tools. The enclosure is designed to be 3D printed at the moment. The material is ABS (Acrylonitrile butadiene styrene) or PETG (Polyethylene terephthalate glycol). We highly recommend using PETG since PETG is UV resistant, high durability while staying in the solid form up to 80$^{o}$C. Thus, PETG is more 3D printing friendly in the lab due to no smell and warping without a 3D printer enclosure.

The small 1.2mm acrylic lens (Fig.4-3) is glued to the front cover (Fig.4-1). Acrylic is UV resistant, low cost, and has good transparency, which makes it suitable for this design. On the back of the back cover (Fig.4-2), there are slots (Fig.4-6) for the zip ties and pipe clamps for the installation at the data logging location. At the bottom of the front cover (Fig.4-1), there are passive and active air inlets (Fig.4-4,5).

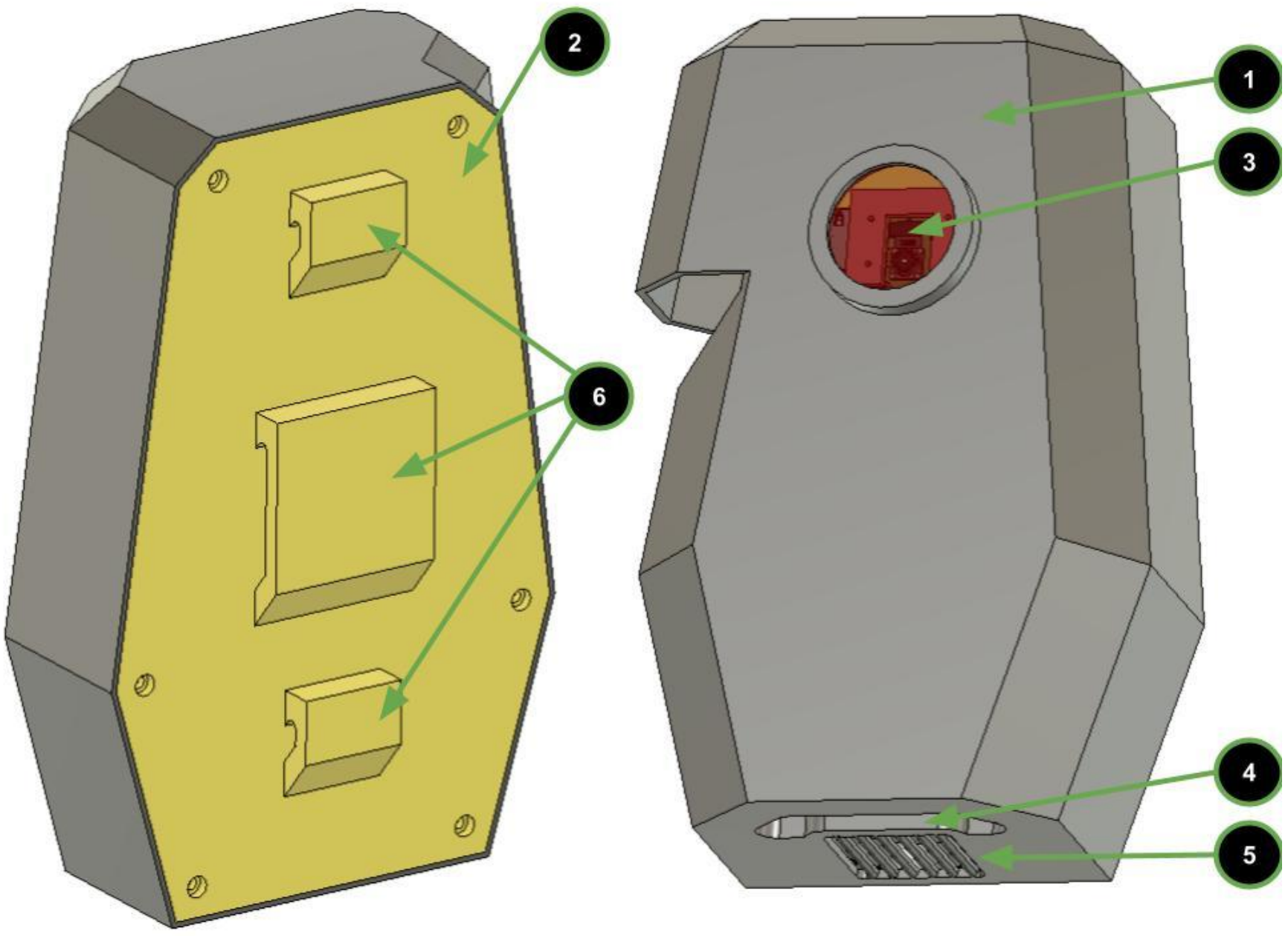

**Fig. 4.** Enclosure (1-Front cover, 2-Back cover, 3-Camera lens, 4-Passive air inlet, 5-Active air inlet, 6-Slots for zip ties and pipe clamps)

PMS 7003, BME 280 and 4040 fan (Fig.5-4,5,6) are mounted on the back cover through snapping clips (Fig.5-1). The Raspberry Pi 3B + and STM-32 Blue Pill development board PCB are mounted on a tray (Fig.5-2) and then snapped to the back cover (Fig.5-1). The tray (Fig.5-2) holds the camera mount (Fig.5-3) with a 8 Mpx camera.

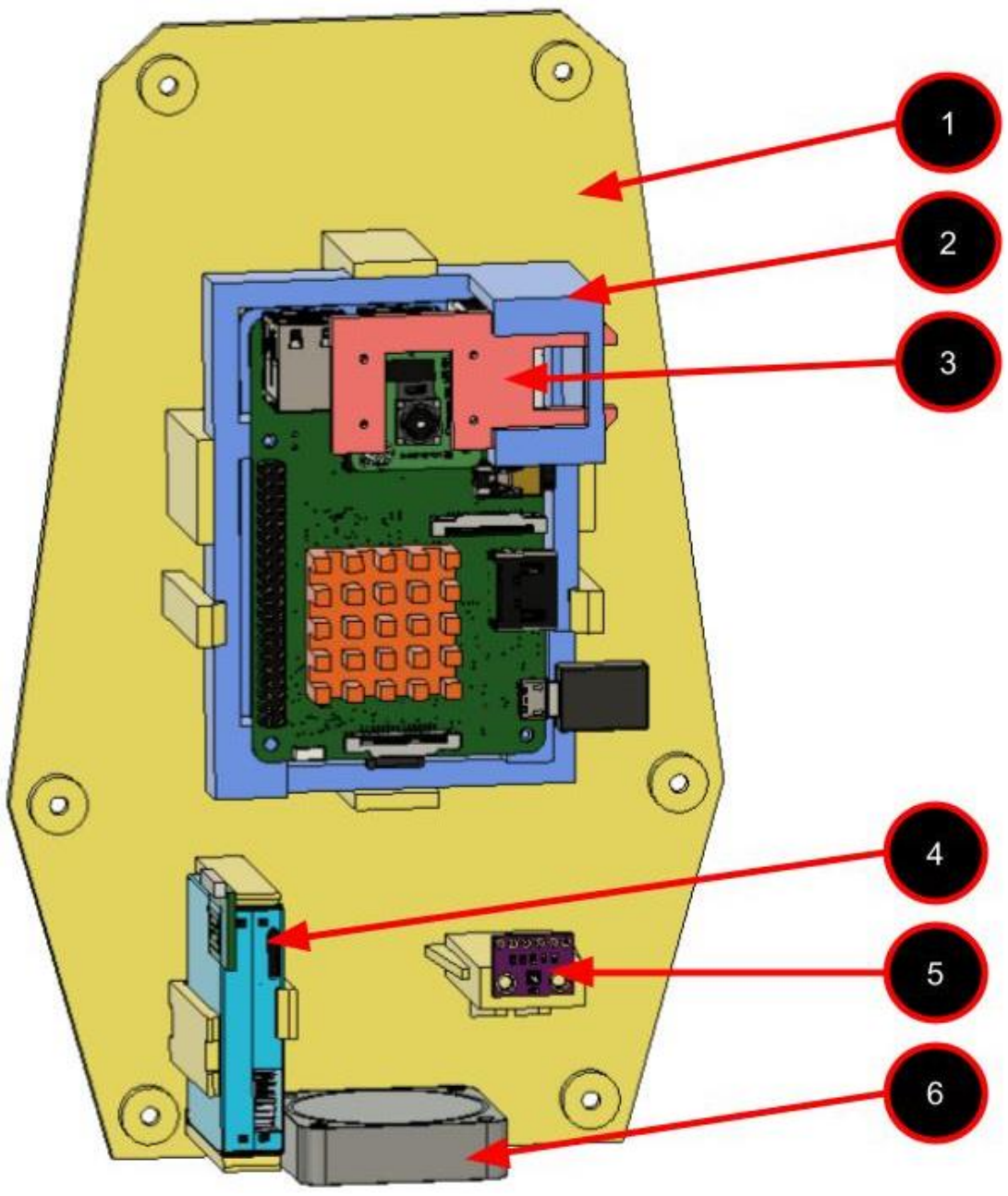

**Fig. 5.** Inside enclosure (1-Back cover, 2-Raspberry Pi 3B+ and STM32 sensor driver PCB mount tray, 3-Camera mount with 8 Mpx camera, 4-PMS7003 particulate matter sensor, 5-BME280 temperature and humidity sensor, 6-Active air intake 4040 fan)

The airflow direction shows that the fresh air travels from the inlet (yellow dashed section) to the outlet (red dashed section), although the airflow velocity direction illustrate there are some turbulence in the middle area close to the Raspberry Pi 3B + heatsink. Since the design can not achieve the laminar flow, dust inside the system will build up over time. That is why the open sensor enclosure design is better for collecting environmental data. The advantage of this closed sensor enclosure is the higher level of weather protection, especially in tropical weather with irregular rain.

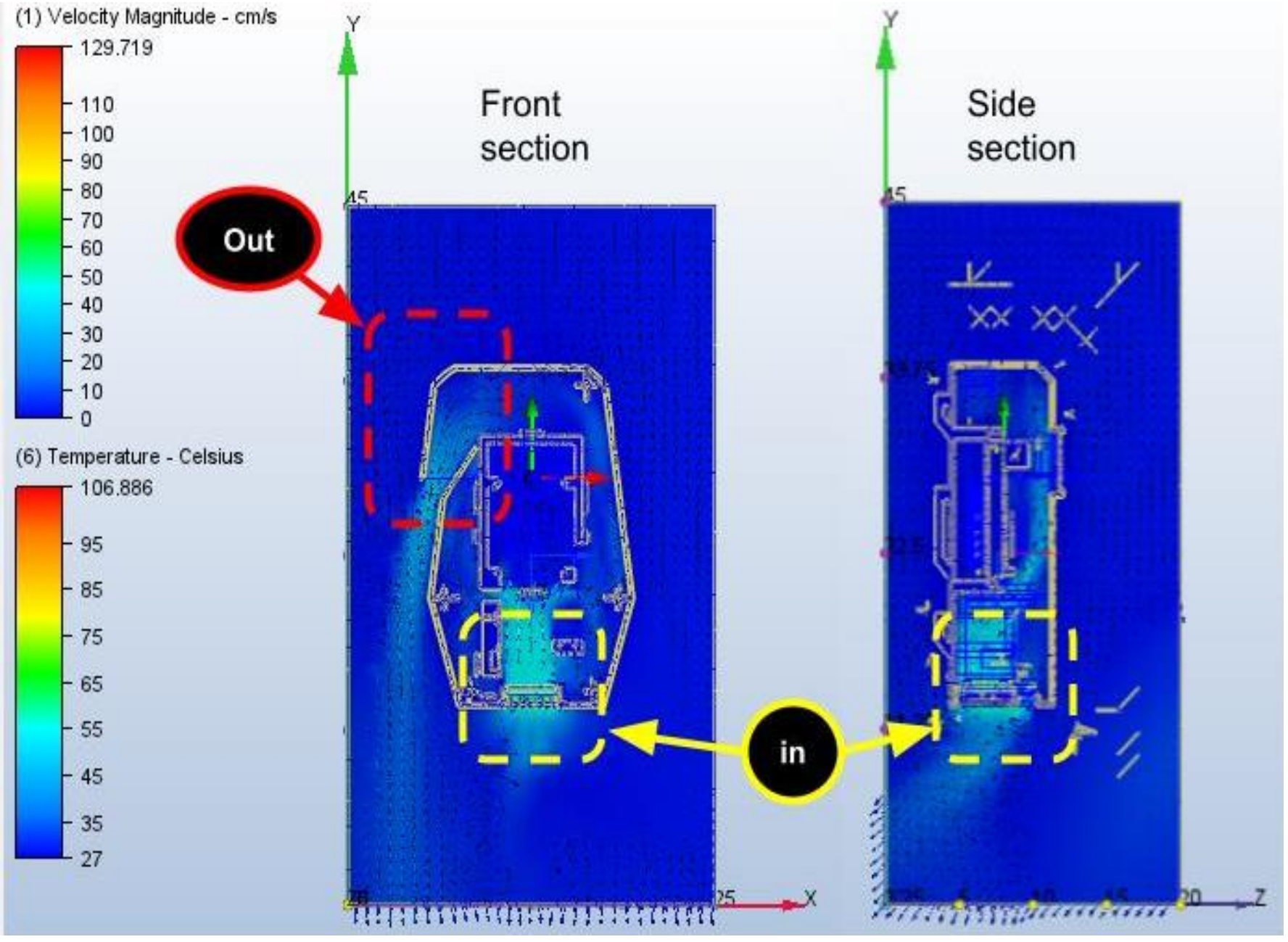

**Fig. 6.** CFD simulation result of heat transfer inside the model (Velocity - cmps scale)

According to ARM Cortex A53 (Raspberry Pi 3B + CPU) specification, it can go up to 105°C, although the Raspberry Pi 3B + operating system is set to throttle at 85°C. In Fig. 7, the hottest temperature area in the center is the CPU heatsink (around 107°C for the worst scenario), with the airflow from the inlet (yellow dashed section) to the outlet (red dashed section), it transfers the heat to the stream and keeps the CPU cool. The nominal temperature in the top area is approximately 45°C to 55°C, while the bottom area temperature (sensor area) is the environment temperature at 27°C. It should keep the environment temperature reading value accurate.

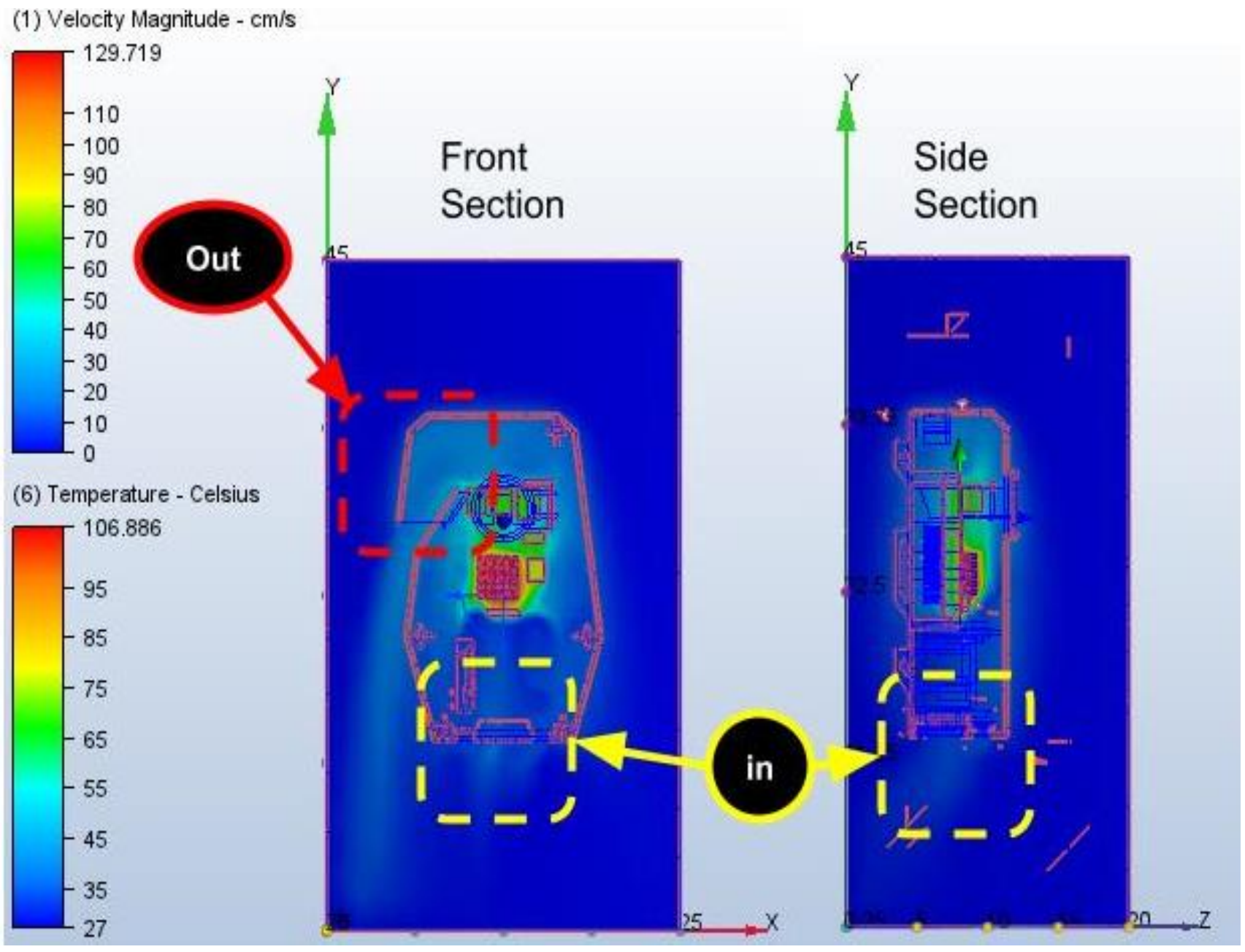

**Fig. 7.** CFD simulation result of heat transfer inside the model (Temperature - Celsius scale)

### 3.5 Vehicle counting and PM2.5 analysis

**Vehicle counting algorithm.** The desire to find the correlation, the relationship between PM2.5 metric and traffic density establishes the need for a vehicle counting algorithm. In order to satisfy the urge, a simple vehicle counting method is implemented, which mainly depends on a certain background subtraction algorithm. The architecture of this method is as follow:

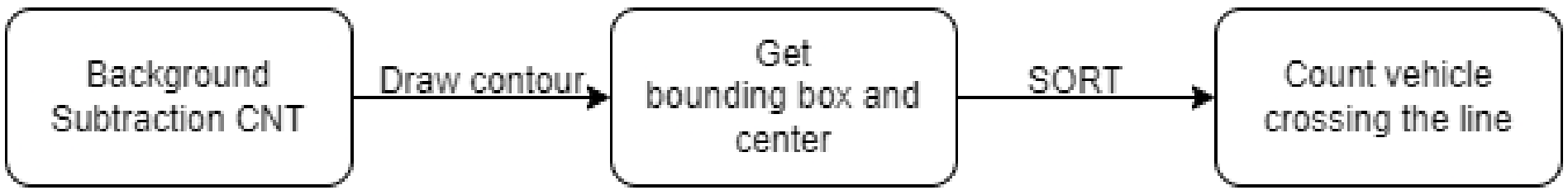

**Fig. 8.** Overview of counting algorithm

To get a foreground object, a background subtraction algorithm called CouNT is used, as this is the fastest option with decent performance for hardware with low spec that is available in OpenCV [8]. After subtracting the background, the foreground will be segmented into different objects through contouring, and then bounding boxes and their centers are achieved.

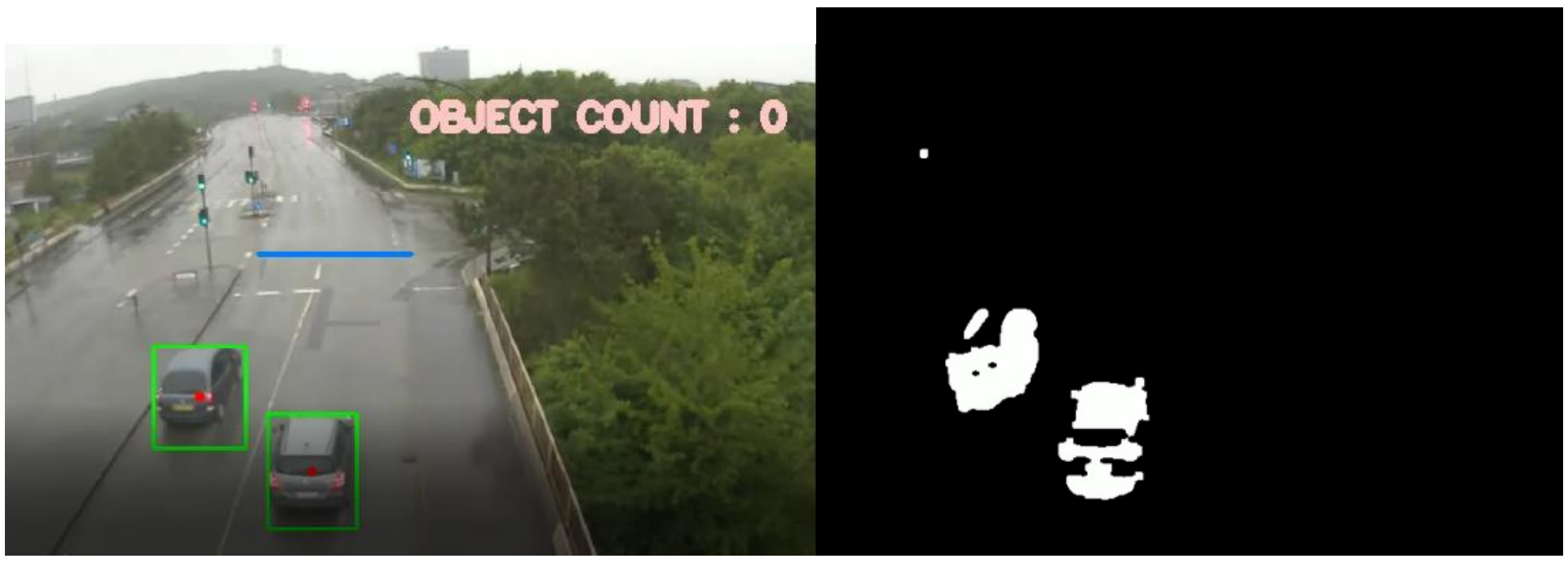

**Fig. 9.** Example of background subtraction. Left figure shows the foreground after subtracting

After the previous steps, each vehicle is counted by line method, where a virtual line is placed on the frame, and when the center of the vehicle crosses this line, the number of vehicles increases by one. This method is enhanced by a tracking algorithm called SORT [9] that utilizes Kalman Filter and Hungarian algorithm for better accuracy.

**PM2.5 analysis.** After being collected, the PM2.5 metric is then analyzed by first being preprocessed and then being visualized via time - series plot and correlation plot. The preprocessing step removes outliers and normalizes the data. It works as follow:

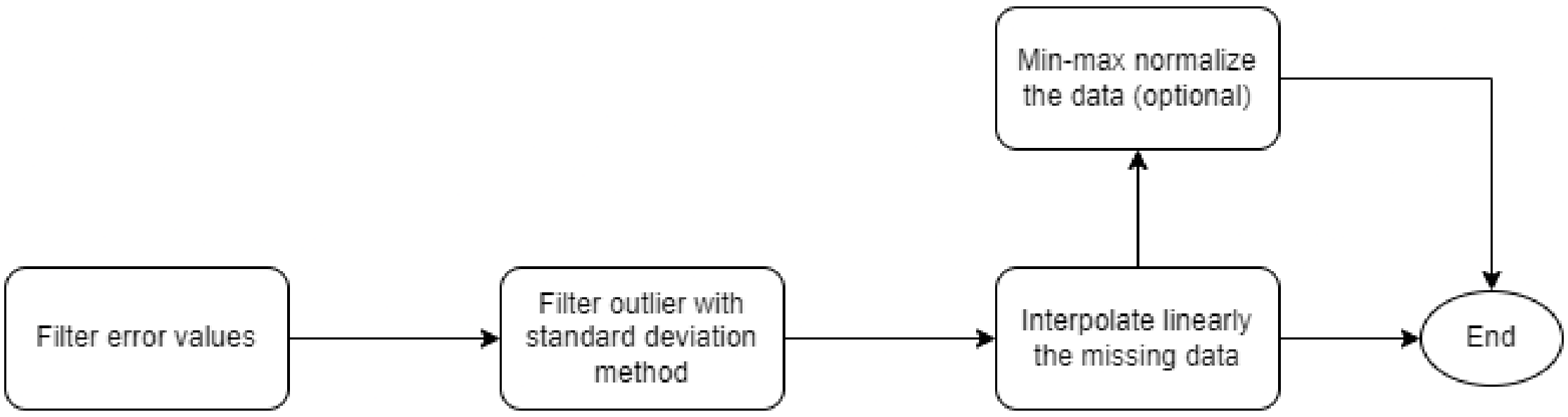

**Fig. 10.** Overview of PM2.5 analysis procedure

In step 1, PM2.5 values which are larger than 20000 (this value is hardware-specific) are filtered out, as these values are errors. In step 2, outliers are removed by using the standard deviation method. In step 3, missing data are linearly interpolated and their sampling rate is altered from second to hour. In step 4, the data is normalized by min-max normalization. Min-max normalization is as follows:

$$x_{scaled} = \frac{x - x_{min}}{x_{max} - x_{min}}$$

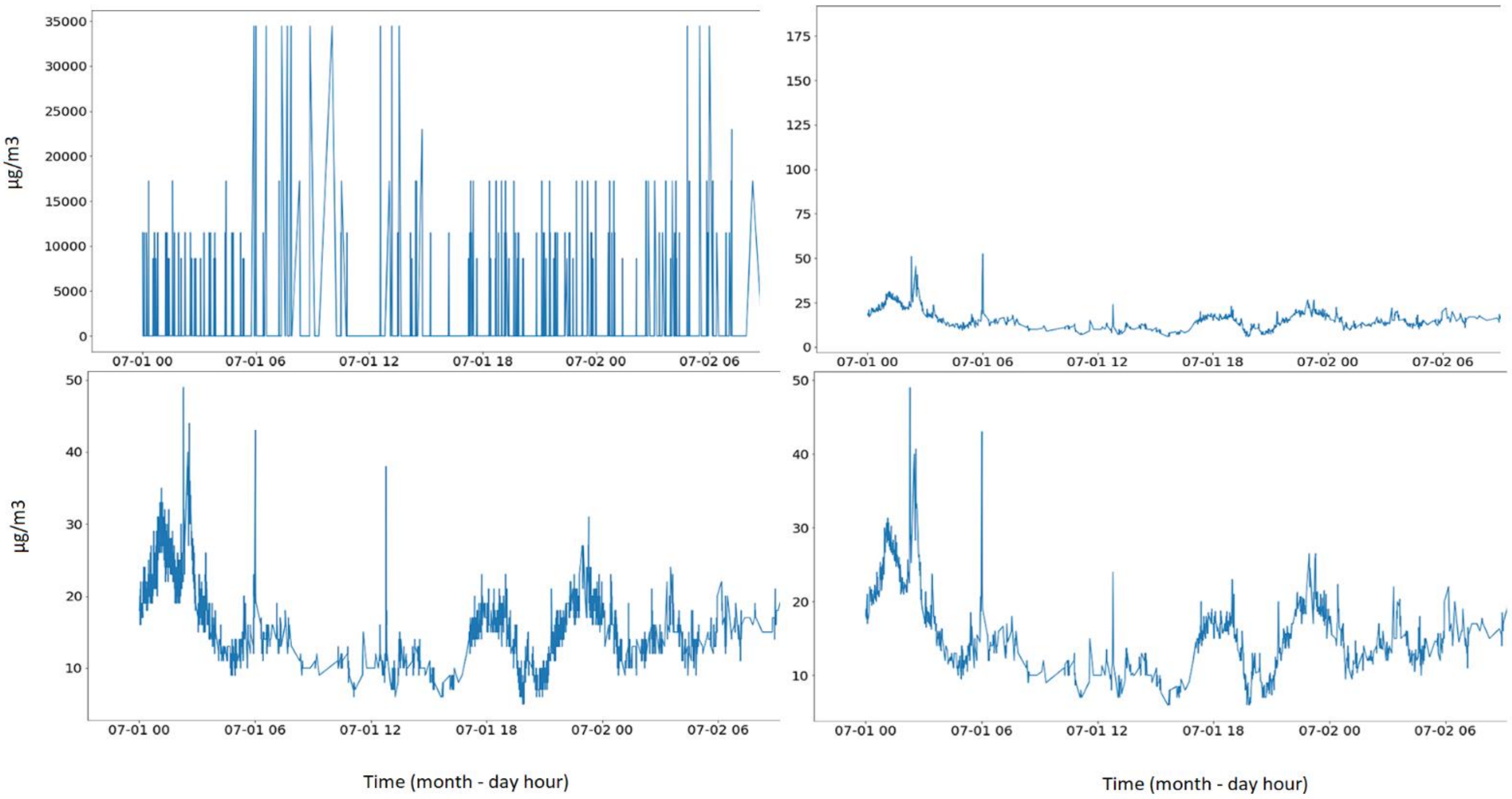

**Fig. 11.** Data before cleaning and after each step (step 1 to 3) (plots are placed in left-right, top-bottom order)

# 4 Results

## 4.1 PMS7003 evaluation and calibration

Because PMS7003 is a low-cost, off-the-shelf sensor, it is recommended that the PMS7003 be calibrated before using [5]. The method is by colocating the PMS7003 with a well-calibrated sensor, usually with superior quality compared to the PMS7003, and then the collected data is evaluated. The goal is to find an offset and scale factor which best correlates the PMS7003 data set with that of the reference. In this paper, we performed the evaluation and calibration process of the PMS7003 with an industrial-grade particle concentration sensor, namely the Dust Trak DRX Aerosol Monitor 8534. Both sensors were located outside to collect PM2.5 dust data for 48 hours. The result will be demonstrated below.

The time series data collected from both sensors are matched along the time axis. Then we performed DTW (Dynamic Time Warping) algorithm and then calculated the 10-minute moving average. The yielded result is demonstrated below.

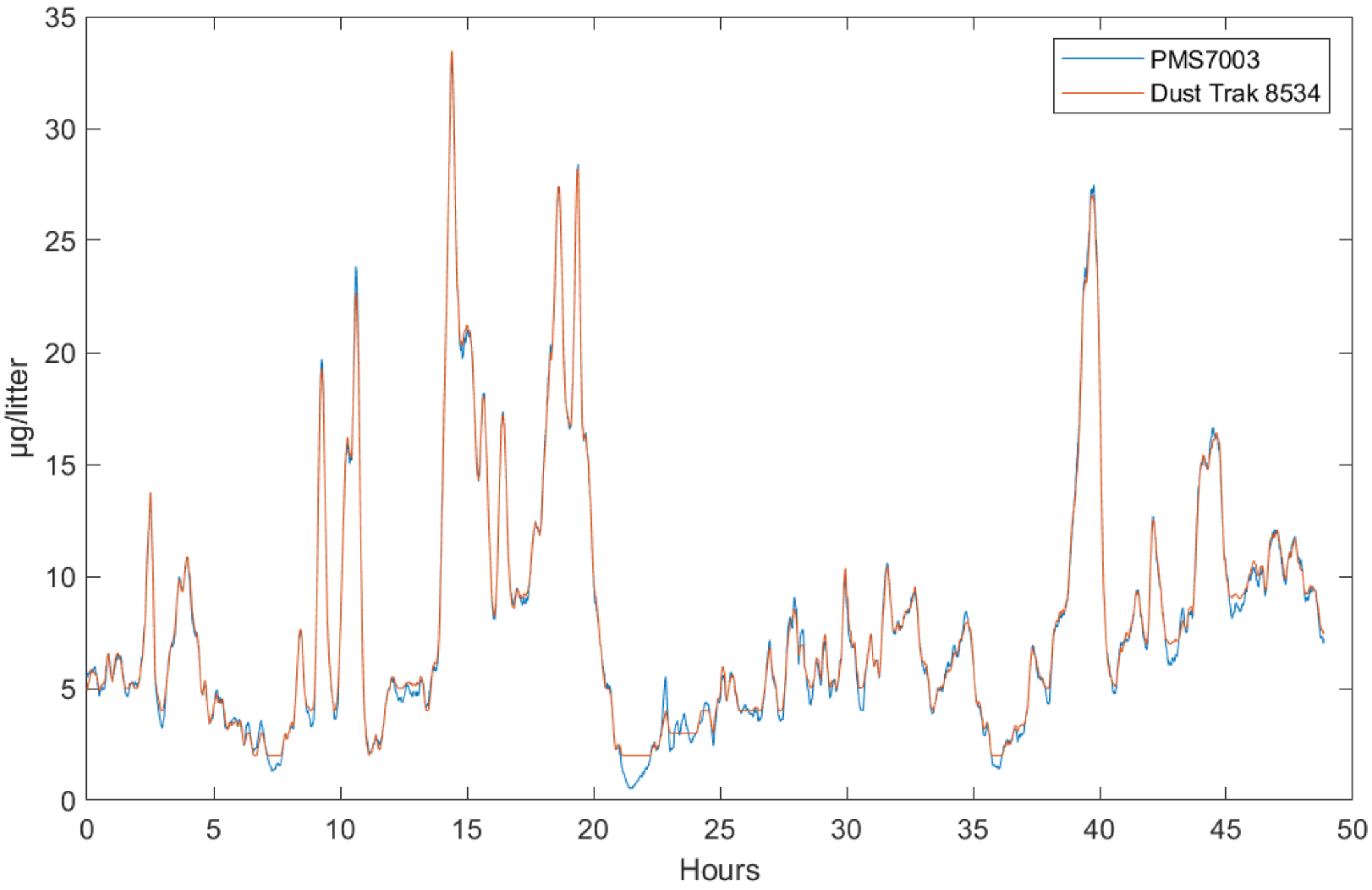

**Fig. 12.** 10-minute moving average plot of PM2.5 time series data from PMS7003 and DustTrak 8534

From the data, a few statistical analysis metrics were calculated.

**Mean Absolute Percentage Error (MAPE).** MAPE metric is used to measure the accuracy of PMS7003 relative to Dust Trak 8534 as a percentage. It is calculated using the average absolute percentage error for each PMS7003 data point minus values from Dust Trak 8534 divided by values from Dust Trak 8534. The result of our data set is 4.95%.

**Root Mean Square Error (RMSE).** RMSE is the square root of the average squared errors. It measures how far away data points from both sensors are, relative to each other. The result of our data set is 0.3254. Because RMSE is dependent on the scale of the data, it should be noted that the range of our data set is 0.5 - 33.5 µg/litter.

**Trend matching score.** We used the Hodrick-Prescott filter to find the trend's cyclical value at each point for each time series. We then compared these values in a piecewise manner. If values from both time series possess the same sign, a trend match is determined. Else, it is a trend mismatch. This score allows us to quantify the general trend match between the two time series. The result of our data set is 72.52%, meaning the trend from PMS7003 data matches that of Dust Trak 8534 approximately 72.52% of the time.

From the analyzed metrics and from qualitative observation of the plotted data, we decided that the specific PMS7003 we used in our system correlated well, relative to the Dust Trak 8534 device. Therefore, we decided that a calibration coefficient is not necessary.

### 4.2 Comparing vehicle density and PM2.5 metric

We compare the PM2.5 metric with the traffic density via time - series plotting. The plot of the result can be found at Fig. A1, appendix.

The 2 lines have some similarities. PM2.5 rises when there are more vehicles, and decreases when there are less vehicles; however, there are some delays, and the pattern between these 2 lines is not as clear as we expected. For example, on 1st of July at around 16 o'clock, the number of vehicles hits a local peak, but the number of PM2.5 hits its peak at around 17 o'clock. Another example would be on 4th of July at 10 o'clock where the vehicle line hits a local peak, while the PM2.5 line hits the local peak at around 12 o'clock. This suggests 2 hypotheses, the first is that the traffic density is not the only factor that affects the PM2.5 metric, but also factors such as weather, temperature, humidity, etc; the second is that vehicle density does not affect air quality immediately, its effect is only visible after a certain amount of delay.

## 5 Limitation

Our storage system is not yet cost-optimized. All the collected data are stored in the "Cool" mode, which facilitates accessibility but generates higher accumulative cost. An algorithm could be developed to automatically move outdated data into "Archive" mode which reduces the cost.

The calibration process of the PMS7003 is still manual, which greatly slows down the building process if more nodes are to be built. An automated process could be developed to automatically analyze and calibrate new PMS7003s.

It is clear that the aforementioned vehicle counting algorithm still has flaws, one of which is that it totally ignores the type of the foreground objects. This counting of course will not show the correct relationship between the number of vehicles and the value of PM2.5 metric, as different types of vehicle will emit different amounts of exhaust fumes. Therefore it is being developed, by being prepended into a second stage called classification, which will help classify the foreground object, thus yield more accurate results.

Although thorough testing has been done, the full system has not been deployed in real traffic for long periods of time. Therefore, the robustness of the system in terms of operation is not determined.

## 6 Conclusion and Outlook

We have integrated the visual and environmental data measurements into one system to simplify installation and management. Although the design shows some weakness in the circulation inside the enclosure, it still meets the need to protect the electronics while logging the data. Although the particular PMS7003 we used for our system did not need calibration, the analyzing process still needs to be performed for every new node built. This will slow down the building process significantly.

In the future, we aim to improve the current design to eliminate the deficiency. Plus, we plan to try another design with exposed sensors while still withstanding the outdoor environment without damaging the sensors and controller boards. A more straightforward calibration process for PMS7003 boosts the fabrication process, which is essential for placing multiple nodes around. Another goal is to optimize this gadget to become an edge device. That means each node can process the raw logged data and save the processed data to Azure Blob storage with less storage.

About the counting algorithm, the current version of it has limitations, as stated above. We are improving it by turning it into a two-stage object detection model. Doing so will allow the algorithm to categorize vehicles and give us more detailed results. From there, we hope to see a more direct cause-effect relationship between vehicle density and PM2.5.

## References

1. Sahanavin, Narut et al. "Relationship between PM10 and PM2.5 levels in high-traffic area determined using path analysis and linear regression." Journal of environmental sciences 69 (2018): 105-114
2. Zhang, H., Srinivasan, R., & Ganesan, V. (2021). Low Cost, Multi-Pollutant Sensing System Using Raspberry Pi for Indoor Air Quality Monitoring. Sustainability (Basel, Switzerland), 13(1), 370–. https://doi.org/10.3390/su13010370
3. Momirski, Lucija Ažman, and Tomaž Berčič. "Southern Inner Ring Road in Ljubljana: 2021 Data Set from Traffic Sensors Installed as Part of the Citizen Science Project WeCount." Data in Brief, vol. 41, 2022, pp. 107878–107878, https://doi.org/10.1016/j.dib.2022.107878.
4. STMicroelectronics, "Medium-density performance line Arm®-based 32-bit MCU with 64 or 128 KB Flash, USB, CAN, 7 timers, 2 ADCs, 9 com. interfaces", STM32F103x8 STM32F103xB datasheet, March 2022.
5. Marek Badura, Piotr Batog, Anetta Drzeniecka-Osiadacz, Piotr Modzel, "Evaluation of Low-Cost Sensors for Ambient PM2.5 Monitoring", Journal of Sensors, vol. 2018, Article ID 5096540, 16 pages, 2018. https://doi.org/10.1155/2018/5096540
6. Raspberry Pi camera V2 documentation page, https://www.raspberrypi.com/documentation/accessories/camera.html
7. Azure Blob Storage pricing page, https://azure.microsoft.com/en-us/pricing/details/storage/blobs/
8. Sagi Zeevi, https://github.com/sagi-z/BackgroundSubtractorCNT (2016)
9. Alex Bewley, Zongyuan Ge, Lionel Ott, Fabio Ramos, Ben Upcroft. Simple object and realtime tracking (2017)

# Appendix

**Fig. A1.** Time-series plot of PM2.5 and number of vehicle crossing

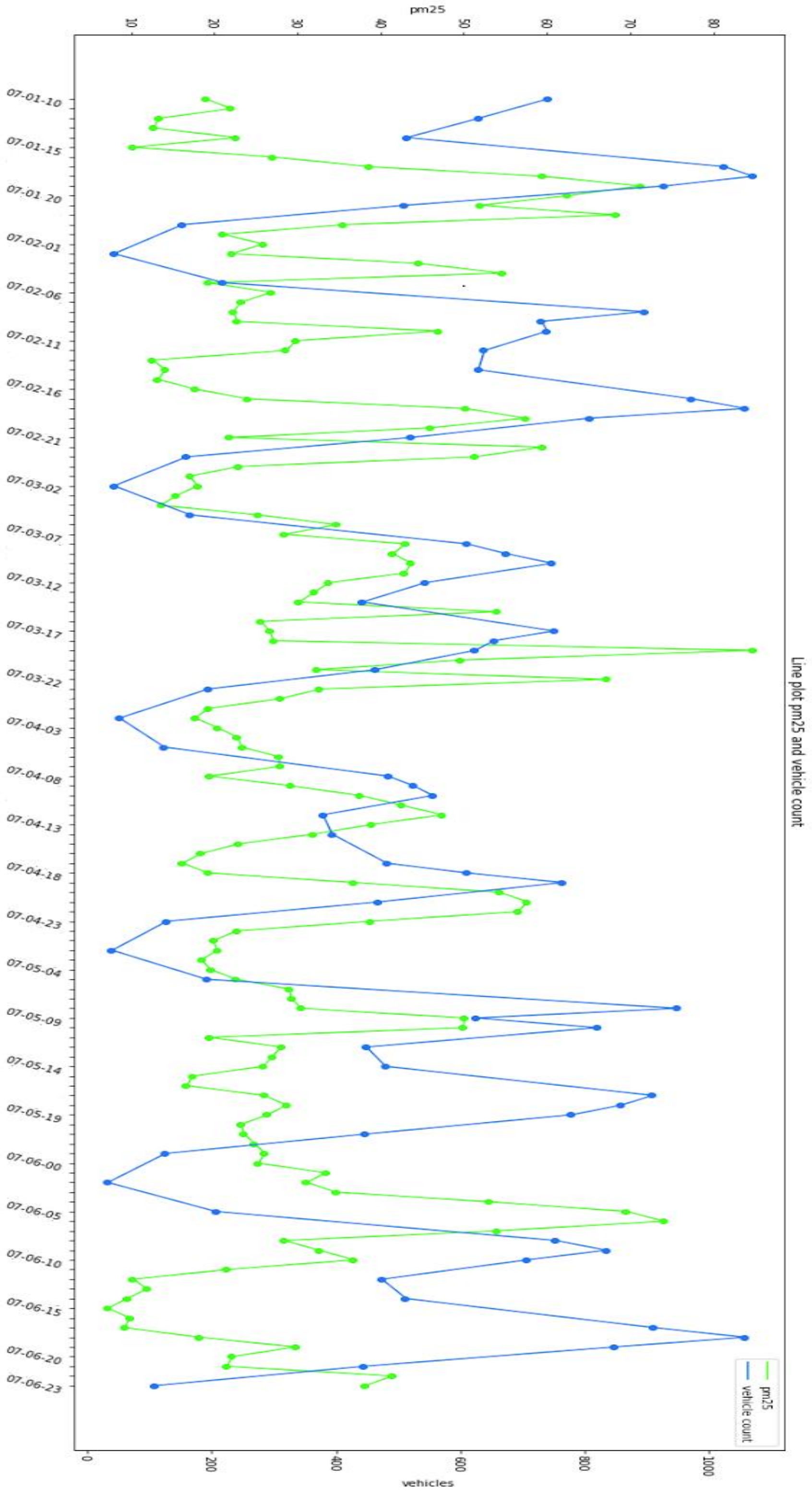